\PassOptionsToPackage{x11names}{xcolor}
\documentclass[12pt]{article}

\usepackage[colorinlistoftodos, textwidth=30mm, shadow]{todonotes}

\usepackage{scicite}
\usepackage{times}
\usepackage{array}
\usepackage{amsmath}
\usepackage{tabularx}
\usepackage{txfonts}
\usepackage{enumerate}
\usepackage{float}
\usepackage{xspace}
\usepackage[x11names]{xcolor}
\usepackage{multirow}
\usepackage{algpseudocode,algorithmicx}
\usepackage{booktabs}
\usepackage{siunitx}
\usepackage{nameref}
\usepackage{makecell}
\usepackage[colorlinks=true,citecolor=red,linkcolor=DarkOrange1]{hyperref}

\usepackage{subcaption}

\usepackage{pdflscape}

\usepackage[T1]{fontenc}
\usepackage[cp1250]{inputenc}

\usepackage{placeins} 

\newenvironment{sciabstract}{%
\begin{quote} \bf}
{\end{quote}}

\title{Identifying Promising Candidate Radiotherapy Protocols via GPU-GA \textit{in-silico}}

\author
{Wojciech Ozimek$^{1}$, Rafa\l{} Bana{\'s}$^{2}$, Pawe\l{} Gora$^{3,\ast}$,\\ Simon D. Angus$^{4}$, Monika J. Piotrowska$^{5}$\\
\\
\normalsize{$^{1}$Ardigen SA, 76 Podole Street, 30-394 Krakow, Poland}\\
\normalsize{ORCID ID: 0000-0003-2588-1353}\\\\

\normalsize{$^{2}$NVIDIA Corporation, al. Chmielna 73, 00-801, 00-001 Warsaw, Poland}\\
\normalsize{ORCID ID: 0000-0002-1811-1206}\\\\

\normalsize{$^{3}$Institute of Informatics, University of Warsaw,}\\
\normalsize{Banacha 2, 02-097 Warsaw, Poland}\\
\normalsize{ORCID ID: 0000-0002-8037-5704}\\\\

\normalsize{$^{4}$Department of Economics and SoDa Laboratories, Monash Business School,}\\
\normalsize{Monash University}\\
\normalsize{Wellington Rd, Clayton, 3800, Australia}\\
\normalsize{ORCID ID: 0000-0001-7095-5054}\\\\

\normalsize{$^{5}$Institute of Applied Mathematics and Mechanics, University of Warsaw}\\
\normalsize{Banacha 2, 02-097 Warsaw, Poland}\\
\normalsize{ORCID ID: 0000-0003-0156-8290}
\\
\\
\normalsize{$^\ast$To whom correspondence should be addressed; E-mail: p.gora@mimuw.edu.pl}
}

\date{}

\begin{document}
\baselineskip24pt
\maketitle

\begin{sciabstract}


Around half of all cancer patients, world-wide, will receive some form of radiotherapy (RT) as part of their treatment. And yet, despite the rapid advance of high-throughput screening to identify successful chemotherapy drug candidates, there is no current analogue for RT protocol screening or discovery at any scale. Here we introduce and demonstrate the application of a high-throughput/\-high-fidelity coupled tumour--irradiation simulation approach, we call `GPU-GA', and apply it to human breast cancer analogue -- EMT6/Ro spheroids. By analysing over 9.5 million candidate protocols, GPU-GA yields significant gains in tumour suppression versus prior state-of-the-art high-fidelity/\-low-throughput computational search under two clinically relevant benchmarks. By extending the search space to hypofractionated areas ($>2$~Gy/day) yet within total dose limits, further tumour suppression of up to 33.7\% compared to state-of-the-art is obtained. GPU-GA could be applied to any cell line with sufficient empirical data, and to many clinically relevant RT considerations.





\end{sciabstract}



\section*{Introduction}

Radiotherapy (RT), and specifically small-dose or fractionated RT, is widely considered to be a cost effective, and potent tool in the treatment of many varieties of cancer with 40 to 50\% of cancer patients receiving some form of RT during their treatment, regardless of the relative prosperity of their context~\cite{Thompson2018}. However, clinical treatment protocols -- the schedule of RT dose size and timing -- are remarkably \emph{singular} in practice despite the wide range of cancers and their pathologies in the human body: `2~Gy per day' being perhaps \emph{the} signature protocol of the last few decades~\cite{Rosenstein:2004vs,ORourke:2009,BoardoftheFacultyofClinicalOncology:2006wq}. Encouragingly, long-running and laborious clinical trials~\cite{Haviland2013,Whelan2010} of slight variations in protocol design (e.g. `3~Gy per day') have shown benefits leading to practice-changing adoptions in clinical settings. As such, given the high cost, and long delays inherent in gathering robust clinical evidence on candidate protocols, there is a pressing need for pre-clinical protocol candidate screening and discovery such that clinical trials examine the most promising alternative candidates. However, away from the tiny number of standard protocols, combinatorial considerations mean that there is an overwhelming number of ways to create a fractionated RT protocol. Consider that, with an alphabet of just five fractional dose sizes, and 10 possible wait-times for the next fraction, a 10 fraction protocol can be constructed in any one of almost $1\times 10^{17}$ ways ($(10\times 5)^{10}$). To make any progress in such a vast search space, any pre-clinical discovery strategy must deliver \emph{high-throughput} screening coupled to \emph{high-fidelity} modelling.




Here we introduce and demonstrate the validation of a high-fidelity, high-performance computational model of human breast cancer analogue, EMT6/Ro spheroids, leveraging state-of-the-art computational approaches on Graphical Processor Unit (GPU) architectures coupled to GA search, an approach we call `GPU-GA', and apply this approach to optimal radiotherapy scheme discovery, at scale. By analysing over 9.5 million candidate protocols, the GPU-GA approach yields improved tumour suppression outcomes of between 4.0\% and 4.5\% versus prior state-of-the-art computational search outcomes under two clinically relevant benchmark constraints, respectively. Significantly, by extending the search space beyond the state-of-the-art constraints, we identify candidate protocols with improved fitness scores of as much as 33.7\% over the best protocols identified in the past work~\cite{Angus2014} under the benchmark conditions studied. By pooling all highly effective protocols discovered in the GPU-GA search activity, we enumerate, and model, the frequency surface of these potent [dose,delay] sub-modules, finding an imputed maximum potency frequency location of 1.4~Gy fraction administered with a 15.5~h delay (after interpolation). Finally, by leveraging the outcomes of the response surface activity, we demonstrate how hand-crafted candidate protocols composed of the most potent sub-modules can be formed and tested, in the framework of computationally-supported treatment design.

Our work extends a previous rather low throughput/high-fidelity methodology~\cite{Angus2014} and builds on the promising field of `computational oncology' (c.f.~\cite{Gerlee2007,Patel2001,Dormann2002,Kansal2000,Kansal2000a,Kempf2015}) which aims to offer a~better understanding of cancer dynamics {\it `in silico'} being the base for improving therapeutic outcomes {\it in vivo}. Importantly, we based on our previous work (c.f.~\cite{Piotrowska:2009gj,Angus:2010mo,Angus:2013id}) where we show that computational cellular automaton (CA)~models of EMT6/Ro spheroid growth can be successfully calibrated to a broad range of experimental data including 18 independent experimental multi-fraction studies~\cite{Otsuka2011,Sugie2006}. Achieving that designed model required  direct modelling of cell metabolism (depending on the local environmental conditions), cell division, DNA damage and repair processes following known biological systems. This provides an extremely promising baseline for the introduction of radiotherapy and/or chemotherapy modules to investigate therapeutic approaches. To this point, of the few authors that have attempted to implement a~computational tumour with a~therapeutic model 
(e.g.~\cite{Zacharaki2004,Stamatakos2006,Kempf2010, Powathil2012,Powathil2013,Kempf2013}) none of them have fully, if at all, calibrated the computational model to experimental data nor applied a~systematic search of the possible treatment protocols. On the other hand, our quasi-3D CA models of avascular tumour growth, built over several years and rigorously calibrated and tested at each stage of development, provide a rich, \emph{high-fidelity} computational model of tumour growth under irradiation.

Central to the high throughput approach, we develop in the present work, is the translation of an existing high-fidelity EMT6/Ro CA model of avascular tumour growth from its parallelised, classical CPU-based {\sc Matlab} language implementation into a parallelised GPU-architecture C++ implementation, delivering c. 717x speed up in candidate protocol evaluation times. As such, we leverage similar approaches in the rapidly emerging field of high-throughput screening (HTS), where a mixture of processor-, big-data-, statistical search/learning tools-, and novel computational architectures are converging to deliver rapid breakthroughs in molecular candidate identification~\cite{Liu2017,Lightbody2019}.

Taken together, we demonstrate the power of GPU-GA computational biology to rapidly and effectively search within a vast solution space for candidate protocols that could be explored in the clinic. Indeed, GPU-GA has much potential for computational oncology research, as it could be applied to any cell line with sufficiently available validation data, and could lead to the rapid identification of radiotherapy schemes to be tried in the clinical setting.

\section*{Results}

\subsection*{Defining the search problem}

Following previous work \cite{Angus2014}, we focus our attention on two clinically relevant benchmark protocols \cite{Rosenstein:2004vs,Bentzen2008, Whelan2002}, both delivering 10~Gy total irradiation in multi-fractions~\cite{Rosenstein:2004vs,ORourke:2009,BoardoftheFacultyofClinicalOncology:2006wq}, over a virtual 5 day period (e.g. Mon-Fri): 
\begin{enumerate}
    \item a two fraction per day, 1.25~Gy per fraction protocol, with 6~h/18~h time delays between doses given at e.g. 9am/3pm, daily --- hereafter `Benchmark I', or BMI,
    \item a one fraction per day, 2~Gy per fraction protocol with doses given every 24~h e.g. at 9am ---  hereafter, `Benchmark II', or BMII.
\end{enumerate}

Let $\mathfrak{P}$ be a set of all protocols $p_i$ composed of a sequence of $k$ pairs of an irradiation dose, $d_i^j \in \mathfrak{D}$ (where $\mathfrak{D}$ is a set of allowed sizes of doses, in Gy, $j\in \{1,\ldots, k\}$), and a time delay  representing wait-times between the previous fraction and the current one, $t_i^j \in \mathfrak{T}$ (where $\mathfrak{T}$ is a set of allowed wait-times, in h), such that  $p_i = ((d_i^1,t_i^1),\dots,(d_i^k,t_i^k))\in \mathfrak{P}$ (for the first given dose, the corresponding wait-time should be interpreted as the time from the beginning of the simulation). Then, according to the definition, BMI and BMII protocols can be represented as $p_{BMI} \equiv ((1.25,0),(1.25,6),(1.25,18),(1.25,6),(1.25,18),(1.25,6),(1.25,18),(1.25,6))$ and $p_{BMII} \equiv ((2.0,0),(2.0,24),(2.0,24),(2.0,24),(2.0,24))$, respectively.

Now, given a function $f$ to determine the fitness of a protocol (see the \nameref{sec:materialsandmethods} section for the definition and the methodology applied to obtain the fitness over multiple synthetic tumours), the general search problem resolves to,
\[
\max_{p_i\in \mathfrak{P}} f(p_i), \quad \text{such that} \quad \sum^k_{m=1} d^m_i \leq 10.
\]
The constraint $\sum^k_{m=1} d^m_i \leq 10$ is added to limit the total sum of irradiation doses by a reasonable amount taking into account negative health consequences~\cite{Rosenstein:2004vs,ORourke:2009,BoardoftheFacultyofClinicalOncology:2006wq}. Note that for protocols with a sum lower than 10~Gy we can always improve (or not worsen, at least) the fitness function, but for $2$ protocols with a similar fitness function, the radiologists may prefer the one with a lower sum of doses, so in the search space we consider also protocols with a sum of doses lower than $10$~Gy.

In \cite{Angus2014}, for BMI, the fractional dose was constrained to $\mathfrak{D}' = \{1.25\}$, and time-delays were explored in 30 minute gradations with a minimum interval of 4~h, up to 26~h, $\mathfrak{T}' = \{4,4.5,\dots,25.5,26\}$, while for BMII $\mathfrak{D}'' = \{2.0\}$ and $\mathfrak{T}'' = \{10,10.5,\dots,31.5,32\}$, meaning that for each benchmark case the delay options considered had a 22~h span available and so $|\mathfrak{T}'|=|\mathfrak{T}''|=45$. It was also assumed that the first dose is always given in the first possible time slot, at the beginning of the simulation. The total dose constraint ($\leq$ 10~Gy) implies that $k \leq 8$ under BMI and $k\leq 5$ under BMII search conditions, respectively. To estimate the number of feasible protocols, under given constrains, we implemented and ran a simple program~\cite{SourceCUDA}, which for BMI and BMII gave as about $3.0\times 10^{11}$  and $4.1\times 10^6$ protocols, respectively, showing the size of the space for considered problem.

Nevertheless, the high-speed throughput gain of the present approach encourages a dramatic increase in the search space for a second phase of experiments, where in addition to searching over $\mathfrak{T}$ we also search over a range of irradiation \emph{doses} from 0.25~Gy up to a given maximal fractional dose, $d^{max}$, i.e. $\mathfrak{D} = \{0.25,0.50,\dots,d^{max}\}$. In the considered research, it was assumed that the minimum considered delay and wait-time is 6~h, which means that up to 20 fractions might be delivered over a 5 day treatment, implying that $k\leq 20$. It was also assumed that the total delivered dose must be lower or equal to 10~Gy. This leads to an enlarged search space size, for (say) $d^{max} = 5.0$~Gy, with at least $1.39 x 10^{15}$ protocols, estimated as follows: first, we easily estimate the number of all protocols with doses summing up to $10$~Gy, without the limit for a single dose equal to $5$~Gy, by realising that such a number is just equal to the number of distributing 40 minimal doses ($0.25$~Gy) in 20 time slots which is $\binom{59}{19}$. Then, we subtract protocols that have a single dose greater than $5$~Gy. If such a dose is equal to $K$~Gy ($K \in \{5.25, 5.5, \ldots, 10\}$, so such a dose would have a number of minimal doses ($0.25$~Gy) from the set $\{21, 22,\ldots, 40\}$), then we have $2\times \binom{59 - 4\times K - 1}{40 - 4\times K}$ protocols in which such a dose is in the first or in the last time slot and $(59 - 4\times K - 1)\times \binom{59 - 4\times K - 2}{40 - 4\times K}$ protocols in which such a dose is in one of the time slots in the middle. Such combinatorial considerations imply that exhaustive, brute-force, search is not applicable under current computational architectures, but instead lend themselves to highly non-linear search strategies such as evolutionary search, as is undertaken here.

\subsection*{Establishing the fidelity of the GPU model}

\begin{figure}[ht]
    \centering
    \includegraphics[width = 0.9\textwidth]{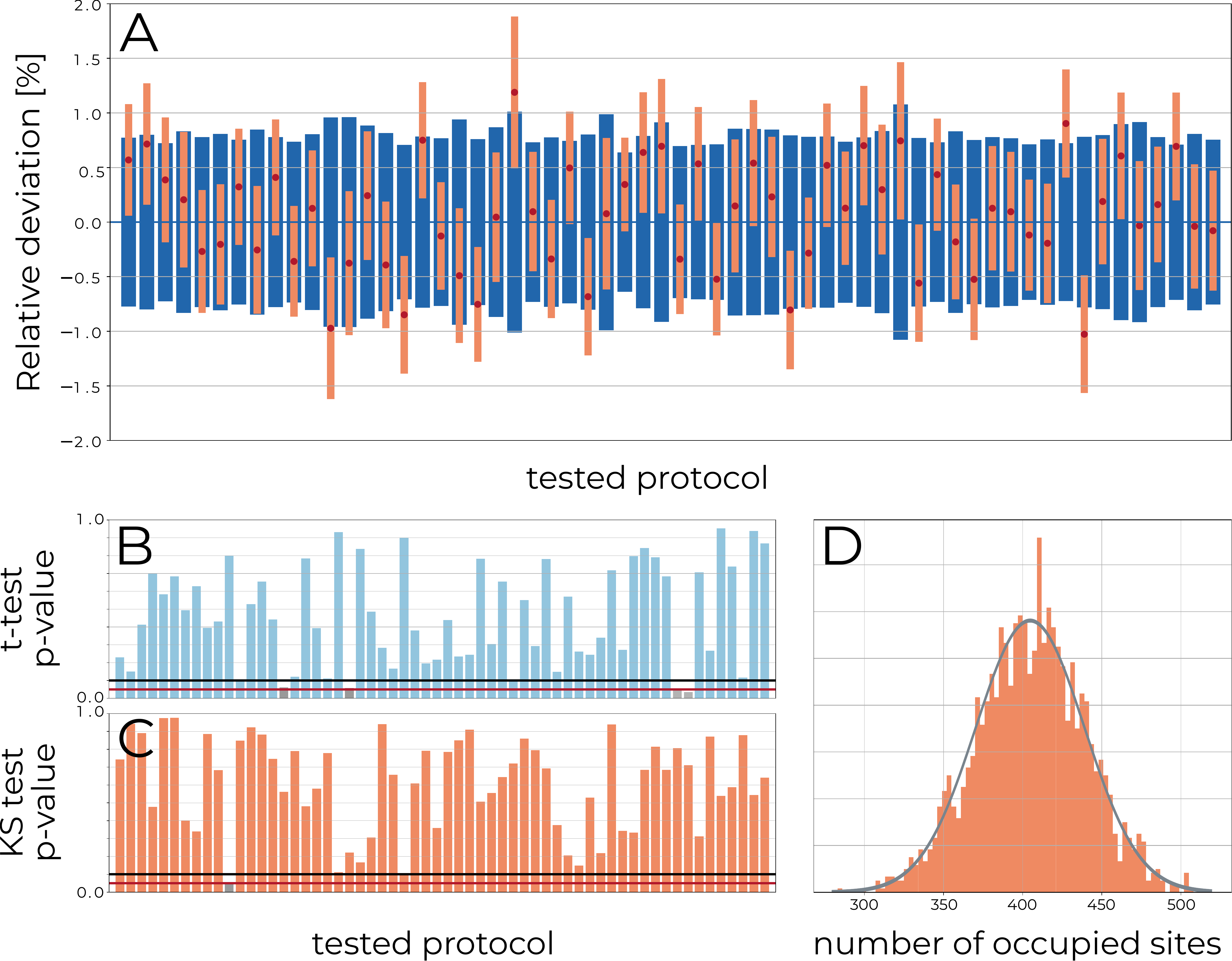}
    \caption{{\bf Examining the fidelity and stability of the GPU implementation of the numerical model.} 60 randomly generated protocols (total dose $\leq$ 10~Gy) were developed and run through 48 and 100 iterations of the CPU and GPU code respectively, against each of 10 tumours in the library for 10 days: (A) shows 95\% confidence intervals for CPU (blue) and GPU (orange) relative deviations from the mean number of occupied sites at the end of each replicate; (B) provides t-test p-values (two sided) comparisons between the CPU and GPU implementations, with 0.1 (black) and 0.05 (red) significance lines shown; (C) provides p-values for Kolmogorov-Smirnov test (KS-test) for the normality of the GPU 100 evaluations over each of 60 protocols, against each of $10$ tumours. In (D)  the histogram arising from 2\,000 GPU evaluations of final occupied sites of a single randomly selected protocol (against a single randomly selected tumour) compared to the normal distribution with the same mean and standard deviation is shown.}\label{fig:CPUvsGPU_normal}
    \end{figure}

First, the fidelity of the C++/GPU implementation of the underlying model to the original parallelised {\sc Matlab} version~\cite{Angus2014} was verified. An intermediate version of the code was written in C++~\cite{SourceCPU}, which accurately represented the dynamics of the original model, but was fast enough to permit baseline comparisons with the developed GPU code. Two forms of verification were conducted -- single step tests (unit tests) and whole run test (integration tests). Using a single tumour from the library, under two different irradiation protocols (a single 20~Gy dose, and a 5 day/2~Gy per day program) the output of the {\sc Matlab} and C++ code were compared and found to exactly match over both unit- and integration- testing regimes~\cite{BachelorThesis}. Next, the GPU implementation~\cite{SourceCUDA} was compared to the intermediate C++ implementation (run on CPU) by comparing the number of occupied sites at the end of an equivalent 5 day treatment period followed by 5 day regrowth period, for 60 randomly generated protocols having total dose less than or equal to 10~Gy. Each protocol was run over 10 test tumours (for details see \nameref{sec:lib} section), for a total of 100 (GPU) or 48 (C++/CPU) replications each. The GPU implementation was found to provide high-fidelity to the base C++ implementation (Fig. \ref{fig:CPUvsGPU_normal}, panel A). Two statistical tests were conducted. First, t-test~\cite{SciPyTTest} was carried out to compare the averages of the results of both (GPU and CPU) implementations (Fig.~\ref{fig:CPUvsGPU_normal}, panel B). Second, Kolmogorov-Smirnov test~\cite{SciPyKSTest} was conducted to investigate the normality of the distribution of the new implementation (Fig.~\ref{fig:CPUvsGPU_normal}, panels C). Finally, 2\,000 evaluations of a single randomly selected protocol were conducted and the distribution of the final number of occupied sites was visualised (see Fig.~\ref{fig:CPUvsGPU_normal}, panel D).

\subsection*{Performance comparison of candidate protocols discovered by GPU-GA}

\subsubsection*{Search under BMI \& BMII}

To accurately assess the fitness of any candidate protocol against a standard benchmark, the fitness methodology (described in \nameref{sec:materialsandmethods} section below) was undertaken for both the clinical benchmarks (BMI and BMII), and the top three state-of-the-art candidate protocols discovered by a previous study~\cite{Angus2014} (i.e. BMI-1, -2, -3, BMII-1, -2, -3). For each protocol, the mean fitness scores and standard deviation was calculated based on fitness function~\eqref{eq:final_fit_fun} over 100 evaluations (random seeds) for each of 10 synthetic tumours from the study library (for details see \nameref{sec:materialsandmethods} section).

{
\begin{table}[htp!]
    \centering
    \caption{
    {\bf Performance comparison of candidate protocol fitness around {\bf BMI}.}
    Mean fitness $f(p)$ under the fitness methodology for each candidate protocol (standard deviation in parenthesis), either as given in the benchmark (BMI), or as discovered by limited (time-delay only) search by \cite{Angus2014} (Panel A); or, as found by this study (Panel B). GA types indicate tournament (Tmnt), top-1 (Top1), or simple (Smpl) selection strategies. `*' indicates values fixed (not searched). Italic and boldface $f$ values indicate minimum and maximum respectively. $\dagger$ significance ${p < 0.001}$, comparing to BMI.}\label{tab:ga_bmi_results}
    \footnotesize 
    \begin{tabular}{lcccccc}
\hline
\hline
Id & $p$ & Total &  Max & Min time & GA & $f(p)$ \\
 & ([Gy], [h])  & dose  & dose & interval & type &   \\
 &  & [Gy] & [Gy] & [h] &  & \\ 
 
 \hline\hline
 \multicolumn{7}{c}{\bf Panel A} \\
 \hline
 
 \multirow{4}{*}{BMI}  & ((1.25, 0) (1.25, 6), & \multirow{4}{*}{10} & \multirow{4}{*}{1.25$^*$} & \multirow{4}{*}{6$^*$ and 18$^*$} &  \multirow{4}{*}{-} &  \\ 
                       & (1.25, 18), (1.25, 6), & & & & & \it 1118.95 \\ 
                       & (1.25, 18), (1.25, 6), & & & & & (40.45) \\ 
                       & (1.25, 18), (1.25, 6))  & & & & &  \\ \hline
\multirow{4}{*}{BMI-1} & ((1.25, 0), (1.25, 18.5), & \multirow{4}{*}{10} & \multirow{4}{*}{1.25$^*$} & \multirow{4}{*}{4} &  \multirow{4}{*}{Tmnt \cite{Angus2014}} &  \\ 
                       &  (1.25, 20), (1.25, 18.5), & & & & & 1138.55 \\ 
                       & (1.25, 17.5), (1.25, 16.5), & & & & & (33.60) \\ 
                       &(1.25, 13), (1.25, 10)) & & & & & \\ \hline
\multirow{4}{*}{BMI-2} & ((1.25, 0), (1.25, 18.5), & \multirow{4}{*}{10} & \multirow{4}{*}{1.25$^*$} & \multirow{4}{*}{4} &  \multirow{4}{*}{Top1 \cite{Angus2014}} &  \\ 
                       & (1.25, 20), (1.25, 18.5), & & & & & 1137.48\\ 
                       & (1.25, 17.5), (1.25, 16.5),  & & & & & (33.53) \\ 
                       & (1.25, 12.5), (1.25, 10)) & & & & & \\ \hline
\multirow{4}{*}{BMI-3} & ((1.25, 0), (1.25, 18.5)  & \multirow{4}{*}{10} & \multirow{4}{*}{1.25$^*$} & \multirow{4}{*}{4} &  \multirow{4}{*}{Tmnt \cite{Angus2014}} & \\ 
                       & (1.25, 20), (1.25, 18.5),  & & & & & 1139.26\\ 
                       & (1.25, 17.5), (1.25, 16.5), & & & & & (33.28)\\ 
                       &  (1.25, 14), (1.25, 10)) & & & & & \\
                       
\hline\hline
\multicolumn{7}{c}{\bf Panel B} \\
 \hline

\multirow{4}{*}{GA1.25-1} & ((1.25, 8.5), (1.25, 18.5),  & \multirow{4}{*}{10} & \multirow{4}{*}{1.25} & \multirow{4}{*}{12} & \multirow{4}{*}{Smpl} &  \\ 
                       & (1.25, 18.5), (1.25, 15.5) & & & & &  \bf 1153.62$\dagger$ \\ 
                       & (1.25, 15.5), (1.25, 13.5) & & & & & (34.56)\\ 
                       & (1.25, 15), (1.25, 14)) & & & & & \\ \hline
\multirow{4}{*}{GA1.25-2} & ((0.75, 1.5), (1.25, 12.5),  & \multirow{4}{*}{10} & \multirow{4}{*}{1.25} & \multirow{4}{*}{12} & \multirow{4}{*}{Smpl} &  \\ 
                       & (1, 19.5), (1.25, 15) & & & & & 1148.38$\dagger$ \\ 
                       & (1.25, 14.5), (1.25, 13.5) & & & & & (34.84) \\ 
                       & (1, 13), (1, 14), (1.25, 13.5)) & & & & & \\ \hline
\multirow{4}{*}{GA1.25-3} & ((1, 0), (1.25, 15.5),  & \multirow{4}{*}{10} & \multirow{4}{*}{1.25} & \multirow{4}{*}{12} & \multirow{4}{*}{Smpl} &  \\ 
                       & (1.25, 15), (1.25, 16) & & & & & 1147.08$\dagger$ \\ 
                       & (1.25, 15), (1, 13.5) & & & & & (33.38) \\ 
                       & (1, 14), (1, 14), (1, 16.5)) & & & & & \\ 
\hline
\hline
    \end{tabular}
\end{table}
}



\begin{figure}[h]
    \centering
    \includegraphics[width=0.9\textwidth]{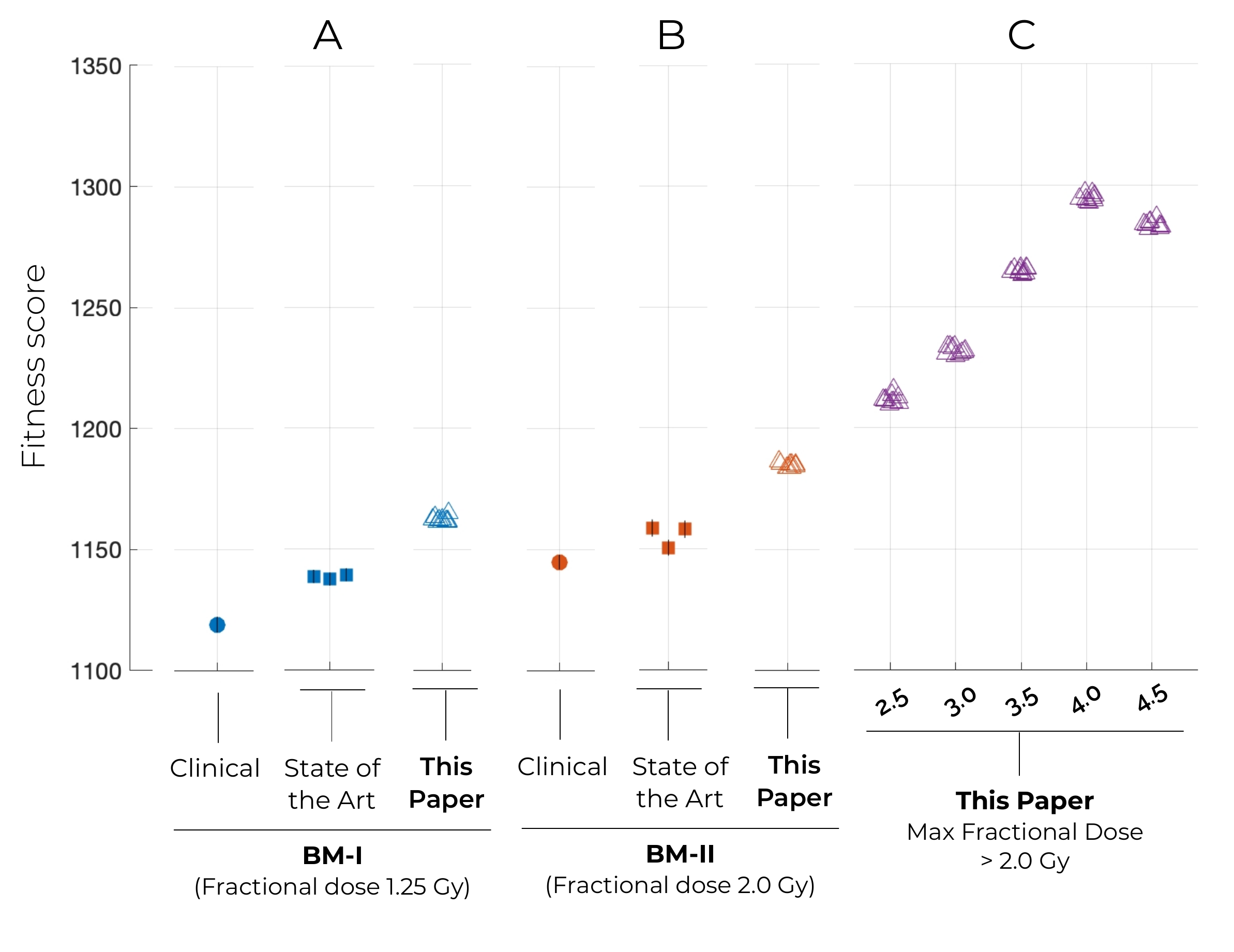}
    \caption{{\bf Performance comparison of candidate protocols versus BMI, BMII and in the multi-parameter case.} Fitness scores (higher values indicate greater tumour suppression) for the clinical (circle markers) and state-of-the-art~\cite{Angus2014} (squares) protocols of prior work, and discovered via GPU-GA in the current study (open triangles), under the constraints of BMI (A), BMII (B) and the multi-parameter case (C). Markers for the GPU-GA represent the top 10 discovered protocols (n.b.: x-position randomly moved to prevent over-plotting). Note: in all cases $\sum^k_{m=1} d^m_i \leq 10$.}\label{fig:benchmark_comparison}
\end{figure}

Panel A of Fig.~\ref{fig:benchmark_comparison} and Panel A of Table~\ref{tab:ga_bmi_results} present the results of the performance comparison between the BMI standard and the previous state-of-the-art~\cite{Angus2014} under the fitness methodology of the present work. In Panel B of Table~\ref{tab:ga_bmi_results} the best performing protocol candidates identified under GPU-GA search are reported where fractional doses of \emph{up to} 1.25~Gy were considered, with the top, average fitness outcome of 1153.62, representing a 4.0\% improvement over the prior BMI state-of-the-art, or 9.1\% gain over the BMI outcome (in terms of tumour suppression, based on formula \eqref{eq:supression}). In all cases, the top three GPU-GA protocols were discovered by a GA search strategy (see \nameref{sec:GA_approach} section for more details on GA methodology).

{
\begin{table}[htp!]
    \centering
    \caption{ 
    {\bf Performance comparison of candidate protocol fitness around {\bf BMII}.}
    Mean fitness $f(p)$ under the fitness methodology for each candidate protocol (standard deviation in parenthesis), either as given in the benchmark (BMII), or as discovered by limited (time-delay only) search by \cite{Angus2014} (Panel A); or, as found by this study (Panel B). GA types indicate top-1 (Top1), top-5 (Top5), tournament (Tmnt), simple (Smpl), or roulette (Rlt) selection strategies. See caption to Table~\ref{tab:ga_bmi_results} for other notes. $\dagger$ significance ${p < 0.001}$, comparing to BMII.}\label{tab:ga_bmii_results}
    \footnotesize
\begin{tabular}{lcccccc}
\hline
Id & $p$   & Total &  Max & Min time & GA &  \\
 & ([Gy], [h])  & dose  & dose & interval & type & $f(p)$ \\
 &  & [Gy] & [Gy] & [h] &  & \\
 \hline\hline
  \multicolumn{7}{c}{\bf Panel A} \\
 \hline
\multirow{3}{*}{BMII} & ((2, 0), (2, 24), & \multirow{3}{*}{10} & \multirow{3}{*}{2$^*$} & \multirow{3}{*}{24$^*$} &  \multirow{3}{*}{-} & \it 1144.83 \\ 
                       & (2, 24), (2, 24), & & & & & (37.69)\\ 
                       & (2, 24)) & & & & & \\ \hline
\multirow{2}{*}{BMII-1} & ((2,0) (2, 15), (2, 22), & \multirow{2}{*}{10} & \multirow{2}{*}{2$^*$} & \multirow{2}{*}{10} &  \multirow{2}{*}{Top5 \cite{Angus2014}} & 1158.54 \\ 
                       & (2, 15.5), (2, 23.5)) & & & & & (43.68) \\\hline

\multirow{2}{*}{BMII-2} &  ((2,0) (2, 15), (2, 19), & \multirow{2}{*}{10} & \multirow{2}{*}{2$^*$} & \multirow{2}{*}{10} &  \multirow{2}{*}{Top5 \cite{Angus2014}} & 1150.39 \\ 
                       & (2, 15.5), (2, 18)) & & & & & (40.56)\\\hline

\multirow{2}{*}{BMII-3} & ((2,0) (2, 15), (2, 23), & \multirow{2}{*}{10} & \multirow{2}{*}{2$^*$} & \multirow{2}{*}{10} &  \multirow{2}{*}{Top1 \cite{Angus2014}} & 1158.03 \\ 
                       & (2, 15.5), (2, 22.5)) & & & & & (44.14) \\

\hline\hline
 \multicolumn{7}{c}{\bf Panel B} \\
 \hline

\multirow{4}{*}{GA2.0-1}   & ((0.25, 17.5), (1.75, 17),  & \multirow{4}{*}{10} & \multirow{4}{*}{2} & \multirow{4}{*}{12} & \multirow{4}{*}{Tmnt} &  \\ 
                       & (1.5, 20.5), (1.5, 15) & & & & & \bf 1173.96$\dagger$ \\ 
                       & (1.75, 16), (1.5, 18.5) & & & & &  (37.80)\\ 
                       & (1.75, 15)) & & & & & \\ \hline
\multirow{4}{*}{GA2.0-2}   & ((0.5, 11), (1.5, 18),  & \multirow{4}{*}{10} & \multirow{4}{*}{2} & \multirow{4}{*}{12} & \multirow{4}{*}{Smpl} & \\ 
                       & (1.75, 20), (1.75, 16.5) & & & & & 1172.55$\dagger$ \\ 
                       & (1.75, 16.5), (1.5, 16.5) & & & & & (38.71) \\ 
                       & (1.25, 16)) & & & & & \\ \hline
\multirow{4}{*}{GA2.0-3}   & ((0.25, 2), (0.25, 15.5),  & \multirow{4}{*}{10} & \multirow{4}{*}{2} & \multirow{4}{*}{12} & \multirow{4}{*}{Rlt} &  \\ 
                       & (1.5, 15.5), (1.5, 19) & & & & & 1172.44$\dagger$ \\ 
                       & (1.5, 15), (1.75, 16.5) & & & & & (36.74) \\ 
                       & (1.75, 16), (1.5, 18.5)) & & & & & \\
\hline
\hline
    \end{tabular}
\end{table}
}



Panel B of Fig.~\ref{fig:benchmark_comparison} and Panels A and B of Table~\ref{tab:ga_bmii_results} give the equivalent results under the BMII environment. Here, fitness gains of 4.5\% and 8.2\% were obtained relative to the prior state-of-the-art and baseline BMII protocol, respectively (according to formula \eqref{eq:supression}). Under BMII settings the GPU-GA strategies with the best performance were found to be even distributed from tournament, simple or roulette selection.

\subsubsection*{Multi-parameter search beyond BMII}

In a second phase of search, the maximum fractional dose, $d^{max}$ was incrementally stepped beyond 2.0~Gy (as in BMII), in 0.5~Gy steps towards 4.5~Gy, whilst keeping the total, 5 day, treatment dose at or below 10~Gy. As noted earlier, by relaxing the constraints on $d^{max}$ the search space increases dramatically, potentially frustrating the GPU-GA technique in discovering more promising protocols. However, results presented in Table~\ref{tab:best_ga_larger_dose_results} and given in Panel C of Fig.~\ref{fig:benchmark_comparison} demonstrate that the GPU-GA technique was highly effective in identifying potent candidate protocols under each set of search constraints provided.

{
\begin{table}[htp!]
    \centering
    \caption{
    {\bf Performance comparison of candidate protocol fitness under multi-parameter search.}
    Mean fitness $f(p)$ under the fitness methodology for each candidate protocol (standard deviation in parenthesis), under varying maximum fractional dose constraints (column 4), holding the total treatment dose below or equal to 10 Gy. GA type indicates roulette (Rlt), tournament (Tmnt), or simple (Smpl) selection strategies. See caption to Table~\ref{tab:ga_bmi_results} for other notes. }\label{tab:best_ga_larger_dose_results}
    \footnotesize
    \begin{tabular}{lcccccc}
\hline
\hline
Id & $p$  & Total &  Max & Min time & GA &  \\
 & ([Gy], [h])  & dose  & dose & interval & type & $f(p)$  \\
 &  & [Gy] & [Gy] & [h] & & \\ \hline\hline

 \multirow{3}{*}{GA2.5-1} & ((0.25, 25.5), (1.75, 26.5),  & \multirow{3}{*}{10} & \multirow{3}{*}{2.5} & \multirow{3}{*}{12} & \multirow{3}{*}{Rlt} & 1193.99  \\ 
                       & (2, 16), (2, 19) & & & & &  (42.37)\\ 
                       & (2, 16.5), (2, 15.5)) & & & & &  \\ \hline
 \multirow{3}{*}{GA2.5-2} & ((2, 44.5), (2, 24.0),  & \multirow{3}{*}{10} & \multirow{3}{*}{2.5} & \multirow{3}{*}{12} & \multirow{3}{*}{Tmnt} & 1192.59  \\ 
                       & (2.25, 17.5), (2, 14.5) & & & & & (45.81) \\ 
                       & (1.75, 14)) & & & & &  \\ \hline\hline
\multirow{3}{*}{GA3.0-1} & ((0.25, 11.5), (3, 42), & \multirow{3}{*}{10} & \multirow{3}{*}{3} & \multirow{3}{*}{12} &  \multirow{3}{*}{Rlt} & 1213.25  \\ 
                       & (2.5, 23.5), (2.25, 17), & & & & & (50.57)\\ 
                       & (2, 18.5)) & & & & & \\ \hline
\multirow{3}{*}{GA3.0-2} & ((0.5, 22), (0.5, 14.5), & \multirow{3}{*}{10} & \multirow{3}{*}{3} & \multirow{3}{*}{12} &  \multirow{3}{*}{Tmnt} & 1212.63  \\ 
                       & (0.25, 17.5), (3, 16), & & & & & (50.92)\\ 
                       & (2.75, 23), (3, 20.5)) & & & & & \\ \hline\hline
\multirow{2}{*}{GA3.5-1} & ((2.5, 60.5), (2.5, 21), & \multirow{2}{*}{10} & \multirow{2}{*}{3.5} & \multirow{2}{*}{12} &  \multirow{2}{*}{Tmnt} & 1232.35   \\ 
                       & (2.5, 17.5), (2.5, 16.5)) & & & & & (60.85)\\ \hline
\multirow{3}{*}{GA3.5-2} & ((0.25, 2.5), (0.75, 57), & \multirow{3}{*}{10} & \multirow{3}{*}{3.5} & \multirow{3}{*}{12} &  \multirow{3}{*}{Smpl} & 1231.71   \\ 
                       & (3, 13.5), (2.75, 26.5), & & & & & (52.88) \\ 
                       & (3.25, 19.5)) & & & & & \\ \hline\hline
\multirow{2}{*}{GA4.0-1} & ((3.75, 75.5), (3.75, 24.5), & \multirow{2}{*}{10} & \multirow{2}{*}{4} & \multirow{2}{*}{12} &  \multirow{2}{*}{Smpl} & \bf 1273.7  \\ 
                       & (2.5, 19)) & & & & & (64.75)\\ \hline
\multirow{2}{*}{GA4.0-2} & ((3.25, 60), (3.5, 24), & \multirow{2}{*}{10} & \multirow{2}{*}{4} & \multirow{2}{*}{12} &  \multirow{2}{*}{Smpl} & 1248.25  \\ 
                       & (3.25, 19.5)) & & & & &(67.53) \\ \hline\hline
\multirow{2}{*}{GA4.5-1} & ((0.25, 53.5), (3.5, 22.5), & \multirow{2}{*}{10} & \multirow{2}{*}{4.5} & \multirow{2}{*}{12} &  \multirow{2}{*}{Tmnt} & 1264.54  \\ 
                       & (2.0, 24.5), (4.25, 15)) & & & & & (63.76) \\\hline
\multirow{2}{*}{GA4.5-2} & ((0.25, 3.5), (3.75, 69), & \multirow{2}{*}{10} & \multirow{2}{*}{4.5} & \multirow{2}{*}{12} &  \multirow{2}{*}{Smpl} & 1264.07    \\ 
                       & (4.25, 26), (1.75, 20)) & & & & & (61.92)\\
\hline
\hline
    \end{tabular}
\end{table}
}


Tumour suppression gains over the best performing GPU-GA candidate from the BMII ($d^{max} = 2.0$~Gy, GA2.0-1 protocol) from 6.1\% ($f(p)=1193.99$, $d^{max}=2.5$~Gy, GA2.5-1 protocol) to 30.6\% ($f(p)=1273.7$, $d^{max}=4$~Gy, GA4.0-1 protocol) were observed in the discovered candidate protocols. 
As the experiment settings gradually moved from $d^{max} = 2.5$~Gy to $d^{max} = 4.0$~Gy, in reference to the best state-of-the-art BMII protocol (BMII-3, Table~\ref{tab:ga_bmii_results} Panel A) gains of 10.4\% (GA2.5-1 protocol) and 33.7\% (GA4.0-1 protocol) were observed. Comparing these GPU-GA results with the baseline clinical BMII protocol (BMII Table~\ref{tab:ga_bmii_results} Panel A), we can observe gains of 13.8\% and 36.3\% respectively. A natural progression can be observed in the fitness outcomes discovered in the 10 best performing candidate protocols of the GPU-GA search in Fig.~\ref{fig:benchmark_comparison} from the $d^{max}$ 2.5~Gy to 4.0~Gy settings, before a decline is observed for the 4.5~Gy settings.

By inspection of the discovered protocols in Table~\ref{tab:best_ga_larger_dose_results} (GA4.5-1 and -2) it can be seen that a high degree of heterogeneity was present in the fractional doses for the 4.5~Gy setting, compared with the more even fractional doses of the GA4.0-1 and -2 candidates. It can be hypothesised that as the fractional dose approaches the total dose allowable, the optimisation surface becomes much more complex with a larger fraction of illegal protocols (i.e. failing the total dose constraint) adjacent to a legal, but well performing protocols.

\subsection*{Hand-crafted high-potential candidates through sub-modular pooling}

\begin{figure}
    \centering
    \includegraphics[width=0.9\textwidth]{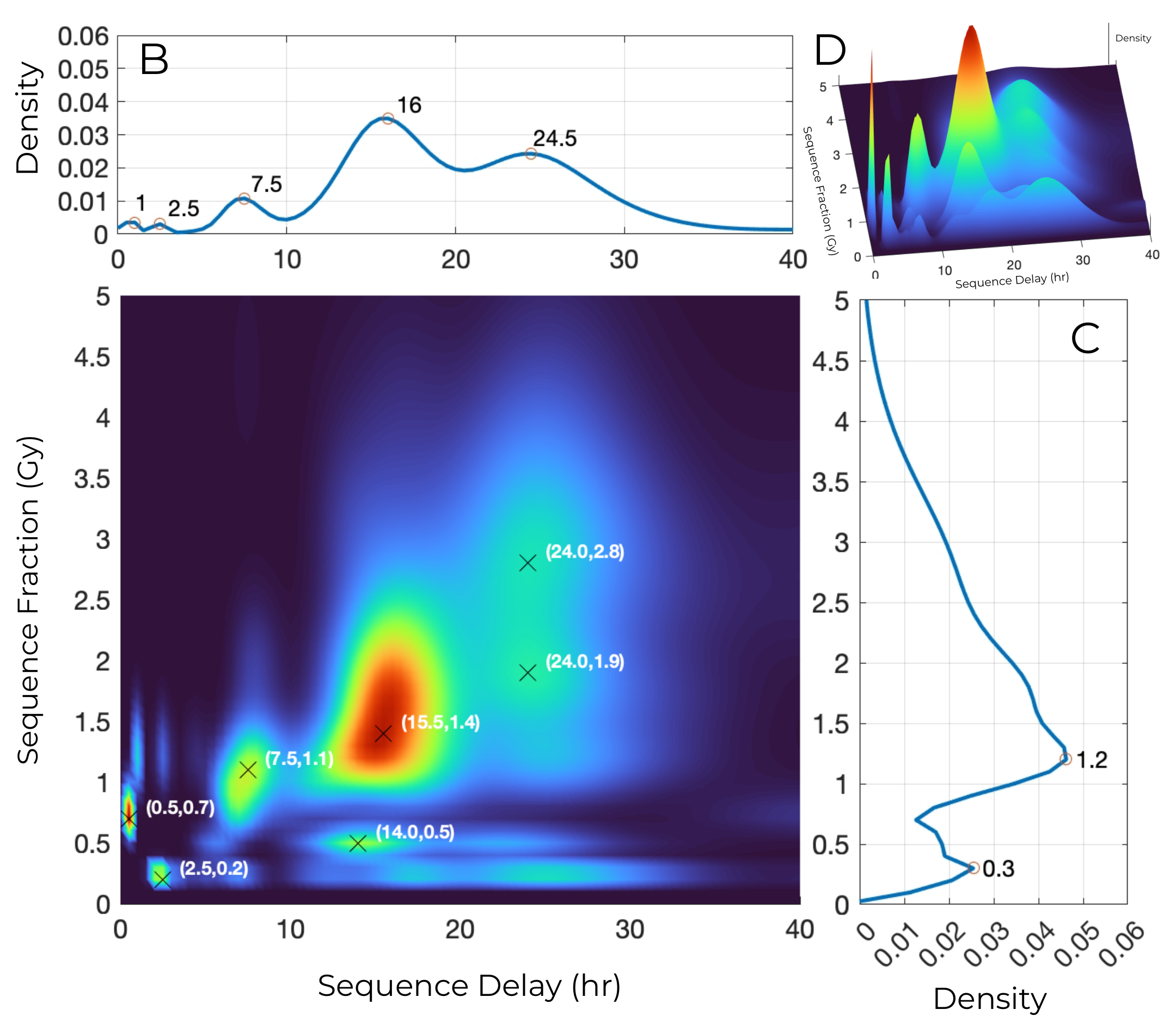}
    \caption{{\bf Pooling density plots of $(t,d)$ pairs from the top 10 protocols from each of the seven $d^{max}$ constraints discovered via the pooled GPU-GA BMI, BMII \& multi-parameter search phases.} All sub-modules are pooled, and a 2-Dimensional kernel density surface is computed (blue indicates low density, red peak density, 3-D visualisation in (D)) (A), with 1-Dimensional smoothed densities provided for sequence delay only (B) and dose fraction only (C) also. The position of prominent peaks are labelled.}\label{fig:density_composite}
\end{figure}

A natural question arises from the GPU-GA search phase described above -- what might be driving the success of the discovered protocols? and, could a \emph{hand-crafted} protocol, informed by the characteristics of the GPU-GA discovered protocols, obtain similar, or even better, performance? Previous work has already hinted at likely reasons for the value of certain protocol dose--delay sub-modules, finding that cell cycle synchronicity might play a role~\cite{Angus2014}. Following this intuition, we take the best 10 performing protocols from each of the seven $d^{max}$ constraints applied across the BMI, BMII and multi-parameter search phases $\{$1.25~Gy, 2.0~Gy, 2.5~Gy, $\dots$, 4.5~Gy$\}$ (the top 2 of which are shown in Table~\ref{tab:best_ga_larger_dose_results}), decompose them into dose--delay sub-modules and then pool all of these together to compute the occurrence density of each sub-module across these high performing protocols (see Fig.~\ref{fig:density_composite}). As can be seen in the figure, a highly frequent peak arises around a fractional dose of 1.4~Gy and delay of 15.5~h, with a sharp, tightly defined peak at 0.7~Gy, 0.5~h also present. The existence of these peaks strongly suggests the presence of particular modalities of multi-fraction protocol design which are potent for the EMT6/Ro cell line, at the stage of growth modelled. If this were not the case, one might expect, for example, either a uniform or Gaussian-like density pattern (somewhat random relationship), or a density pattern matching a particular heuristic (e.g. dividing the given $d^{max}$ by a given number of fractional doses, over 120~h).

{
\begin{table}[htp!]
    \centering
    \caption{
    {\bf Performance comparison of hand-crafted protocols under computationally-supported treatment design.}
    Mean fitness $f(p)$ under the fitness methodology for each candidate protocol tested (standard deviation in parenthesis), where protocol design was inspired by the results presented in Fig.~\ref{fig:density_composite}: HC-1 repeats the (1.4,15.5) pair seven times; HC-2 is the same as HC-1, but replaces the first fraction with the (0.7,0.5) pair; HC-3 replicates HC-2, but adds a (0.7,0.5) pair to the end; and HC-4 is a dummy protocol which aims to tightly fit a maximum fractional dose, 7-repeat protocol into a 120h (5 day) period.}\label{tab:manual_results}
    \footnotesize
    \begin{tabular}{lccc}
\hline
\hline
Id & $p$ & Total dose &  $f(p)$ \\
 & ([Gy], [h]) & [Gy] & (sd)  \\
 \hline
 \hline
 \multirow{2}{*}{HC 1} &  \multirow{2}{*}{$((1.4, 15.5) \dots (1.4, 15.5))$} & \multirow{2}{*}{9.8} & 1147.05 \\ 
                       &  & & (36.31)\\ 
                       \hline
 \multirow{2}{*}{HC 2} &   \multirow{2}{*}{$((0.7, 0.5),(1.4, 15.5) \dots (1.4, 15.5))$}   & \multirow{2}{*}{9.1} & 1123.14 \\ 
                       & & & (37.37)\\ 
                       \hline
 \multirow{2}{*}{HC 3} &  \multirow{2}{*}{$((0.7,0.5) ,(1.4, 15.5) \dots (1.4, 15.5),(0.7, 0.5))$}  & \multirow{2}{*}{9.8} & 1148.79\\ 
                       &  & &  (38.29)\\ 
                       \hline
 \multirow{2}{*}{HC 4} &  \multirow{2}{*}{$((1.42, 17.1) \dots (1.42, 17.1))$}  & \multirow{2}{*}{9.94} & \bf 1157.83\\ 
                       & & & (35.96)\\ 
                       \hline
                       \hline
    \end{tabular}
\end{table}
}

Taken together, a variety of hand-crafted protocols were developed to examine whether a protocol which repeats the high-frequency sub-module, or modules, would be as potent as candidate protocols discovered by GPU-GA (see Table~\ref{tab:manual_results}). Three protocols were developed on the basis of the (1.4~Gy, 15.5~h) sub-module peak: HC-1 repeats the module 7 times; protocols HC-2 and HC-3 replaces the first fraction with the (0.7~Gy, 0.5~h) module (HC-2, HC-3) and then additional adds the same to the end of the protocol (HC-3). Finally, HC-4 was also developed with a simple heuristic based on the findings of the sub-module density surface, namely to adjust the timing slightly from HC-1 so as to deploy uniform dose fractions evenly across the 120~h treatment window but sill close to the main density peak, reaching as close as possible to the 10~Gy total dose limit.

With a maximum dose of 1.4~Gy, the hand-crafted protocols can be compared to previous results lying between BMI (1.25~Gy) and BMII (2.0~Gy) constraints. There, GA1.25-1 (Table~\ref{tab:ga_bmi_results}) with fitness 1153.62 and GA2.0-1 (Table~\ref{tab:ga_bmii_results}) with fitness 1173.96 can be compared to the performance of the hand-crafted protocols HC-1 and HC-2 with fitness of 1147.05 and 1148.79, respectively. Linear interpolation between the performance of GA1.25-1 and GA2.0-1 would suggest that a fitness of 1157.7 would be an approximate ceiling for GPU-GA discovery at 1.4~Gy per fraction. Of the hand-crafted protocols, HC-1 and HC-3 reach close to this level, but the heuristically guided HC-4 achieves an almost identical, theorised GPU-GA like fitness ($f(p) = 1157.83$). These results lend support to the potential of computationally-supported treatment design to efficiently locate candidate protocols for future examination in the clinical setting.

\section*{Discussion}


We have shown the development and application of a high-throughput, high-fidelity, computational screening approach to therapeutic irradiation design for EMT6/Ro spheroids, GPU-GA, which effectively combines state-of-the-art computational architectures with highly non-linear search technology to discover high potential candidates for further study. Through careful calibration of the underlying model, and then translation to GPU articulated form, high-fidelity irradiation protocol evaluations were achieved with c. 717x speed-up relative to the prior, high-fidelity, low-throughput, state-of-the-art~\cite{Angus2014}. The GPU-GA strategy thus enabled the facile exploration of the immense search space produced by even discretised irradiation protocol composition over a multi-day treatment window.

By searching near two clinically relevant benchmarks, the GPU-GA strategy yielded significant tumour suppression gains over both simulated clinical benchmark outcomes and prior, throughput-limited GA search. Beyond the clinical benchmarks, relaxed search constraints saw even larger potential gains. Furthermore, by pooling modular components of the most successful discovered candidates, simple, heuristic-led candidate protocols could be built and tested, yielding close to hypothetical optimal results, even though GPU-GA search was not conducted specifically near the hand-crafted protocol location. 

Our results thus contribute to, and advance, the emerging consensus from theoretical and simulation studies which have identified significant opportunities for treatment efficacy gains beyond the strictures of received, clinical practice. Be it via simulations of ordinary differential equations~\cite{Alfonso2020}, non-linear or linear surrogate modelling~\cite{Montaseri2020}, or agent-based \emph{in silico} modelling~\cite{Kempf2015}, studies that examine deviations from standard `2~Gy per (week) day' treatment modalities have shown that shorter total treatment regimes, hypofractionation, or harnessing immune-response dynamics with immunotherapy drug targeting each present opportunities for treatment optimisation. Whilst none of these related works present high-fidelity model \emph{validation} to a given cell-line, as has been achieved in the current work, all of them present the same conclusion despite their independent and differing model assumptions and approaches, namely that opportunities exist for treatment advance relative to the standard RT approach, and that these opportunities can be explored economically via \emph{in-silico} means.

Our results also align with the findings of the limited clinical experimentation in protocol design that has been conducted in recent years. Notable among these is that of the UK Standardisation of Breast Radiotherapy (START) trials conducted over 1999 and 2001 (START-A $n=2236$, START-B $n=2215$)~\cite{Bentzen2008, Whelan2002}. Ten year follow up studies of the cohorts confirmed that shorter, `more convenient', hypofractionated (RT fractions $> 2.0$~Gy) were no less effective than the world standard longer, 2.0~Gy RT fraction protocols, and were thus adopted by UK cancer treatment centres~\cite{Whelan2010, Haviland2013}. In our multi-parameter search phase, all of which was conducted within a fixed total dose hypofractionated setting, significant, and increasing, potential gains were discovered from 2.5~Gy to 4.0~Gy RT fractions (Fig.~\ref{tab:best_ga_larger_dose_results}). Yet, interestingly, the GPU-GA protocols identified did not follow a regular, 24~h `daily' schedule as the clinical trials considered. Given the success of the computationally-supported protocol design (Table~\ref{tab:manual_results}), together with the clinical findings, a reasonable proposal for next-phase testing would be a hypofractionated, sub-24~h regular, protocol fashioned along the lines of HC-4.

With the benefit of high-throughput, high-fidelity numerical screening that GPU-GA affords, several promising lines of inquiry are now open. First, whilst the present study is built on the EMT6/Ro mammary mouse cell line, GPU-GA could effectively be extended to any cell-line where sufficient \emph{in vitro} empirical data are available. Given the widespread use of the rigid, accepted protocols in the clinical setting across cell lines~\cite{Thompson2018} there seems a high likelihood that similar potential tumour suppression gains will be achieved in other cell lines. Second, and of particular interest for clinical settings of low-compliance due to a variety of factors, GPU-GA could be used to explore the robustness of a given protocol under perturbation. For example, what is the relative impact of a single missed fraction? what if the fraction is missed early, or late in the treatment period? can down-stream impacts of protocol non-compliance be mitigated by adapting the later sequence of fractions? in what way? Whilst these questions are easy to pose, rigorous clinical exploration of these variations would be prohibitively expensive, and time-consuming, yet these are simply alternative specifications to run under GPU-GA and would be entirely feasible for future studies to tackle. Third, and in line with recent calls for more preclinical and clinical trials to `unravel' the best approach to combined RT and immunotherapy scheduling~\cite{Wang2018} GPU-GA could be extended by the development of a combined RT---immunotherapy module to enable rapid prototyping, screening and discovery of candidate joint therapy schedules to be taken to the next phase.

Of course, the GPU-GA strategy outlined herein is not without limitations. First, high-fidelity model calibration depends strongly on the empirical evidence available. Not all cell lines have been studied so rigorously as EMT6/Ro, and, the calibration of the underlying mechanisms of the present model was only achieved under a relatively small (but clinically relevant) spheroidal scale. The dynamics of other cell lines, or growth dynamics under vascularisation will differ substantially and must be organically modelled prior to GPU-GA implementation. Second, the path from candidate protocol discovery to clinical translation will require all of the usual caveats, and lead to some additional challenges. For instance, the optimal fractional delay of the present work, in line with the findings of previous findings~\cite{Angus2014}, of around 15-17~h for EMT6/Ro cells, would translate to RT fractions being administered at variable times through the 24~h period over a treatment regime (e.g. 9~am, 1~am, 5~pm, 9~am, ...). Clinical implementation would require a substantial change to standard day-light hour practices in clinical settings, not to mention patients may require significant support to meet the timetable. Given the complex nature of human systems, irregular RT fractions could also see interactions with other circadian-rhythms of human biology. And, larger fractions, delivered as part of an irregular, quasi-optimal treatment program, may have deleterious side-effects, relative to regular, low-dose fractions. These considerations provide fertile ground for future modelling and discovery across a range of scales.

\section*{Materials and Methods}
\label{sec:materialsandmethods}

\subsection*{Underlying model: source, validation \& calibration}

The high throughput GPU-GA methodology introduced and demonstrated in this study implements exactly a previously published low-throughput, high-fidelity computational model of avascular EMT6/Ro spheroid growth~~\cite{Piotrowska:2009gj, Angus:2010mo, Angus:2013id, Angus2014}. The most recent version of the model~\cite{Angus2014} was re-implemented in C++ \cite{SourceCPU} and then prepared for GPU architectures~\cite{SourceCUDA} in this study. The model takes into account several biologically important components of tumour growth and cellular response to irradiation such as: the diffusion of nutrients (oxygen, glucose) and metabolic waists; cell cycles with distinguishable cell phases;  differentiated cellular metabolism (aerobic or anaerobic, proliferating or quiescent) which is responsive to the cell's environmental nutrient concentrations and local pH. The model was developed to replicate the biological dynamics of a specific mouse cancer cell line (EMT6/Ro), a widely used analogue of human breast cancer which also has abundant experimental studies available. Moreover, the model was calibrated at each step of development, including its response to 18 independent multi-fraction irradiation protocols tested in the laboratory~\cite{Otsuka2011,Sugie2006}. The calibration and validation of the model's performance was accomplished with reference to a~broad spectrum of tumour characteristics including: number of cells; saturation size; tumour volume; tumour doubling time; thickness of the proliferating rim; cell phase population fraction; onset and progression of necrosis; and effective dose induced by multi-fraction irradiation. For details, see~\cite{Angus2014} and Supplementary Information within.



\subsection*{The tumour case library}\label{sec:lib}

Due to the stochastic nature of the tumour growth and radiation response model, we follow the approach presented in~\cite{Angus2014} and perform evaluation of all considered protocols across a multi-tumour case library. The case library consists of ten 10-day old tumours developed {\it in-silico} as described in~\cite{Angus:2013id} from an initial seed population of 200 cancer cells (10 cellular automata lattice grid sites) placed in a well-mixed replenished substrate (glucose concentration: 5.5~mM, oxygen concentration: 0.28~mM, with pH level maintained at 7.4). The tumours were then grown {\it in silico} for 10 days without any radiation interference. For the full characterisation of the developed tumour library we refer the reader to~\cite[Supp. Inf., Table ~S2]{Angus2014}. At the end of the 10-day growth period, the state of each tumour was saved, and stored in the case library for the future use.

\subsection*{GPU implementation: acceleration, fidelity \& normality} 

The key feature of the present study is to articulate the original EMT6/Ro {\sc Matlab} model implementation~\cite{Angus2014} to a new general-purpose C++/GPU implementation to obtain a step-change in protocol evaluation. Applying general purpose GPU programming to obtain high-throughput numerical simulation of candidate protocols was considered highly promising due to the parallelizable nature of cellular automata computation and the fact that evaluation of a radiation protocol requires multiple independent simulation runs. The architectural idea of the new implementation was to use leverage massive parallelism of the GPUs not only to parallelize the processing of single cells within a simulation but also to concurrently run a batch of simulations. With the new GPU implementation~\cite{SourceCUDA}, an average speed of 80\,000 simulations per hour was achieved using 4 NVIDIA V100 GPUs. This compares to about 111 simulations per hour using 13 10-core CPUs as reported in~\cite[Supp. Inf., Numerical Implementation: Details]{Angus2014}, giving c. 717 improvement.

To ensure fidelity of the new GPU implementation~\cite{SourceCUDA}, an intermediate version of the code written in standard C++ was used, which accurately copied the behaviour of the original {\sc Matlab} simulation but was fast enough to provide a convenient baseline to compare the GPU implementation with~\cite{SourceCPU}.  Given the importance of stochastic dynamics in the underlying model, the C++ implementation required replicating exactly the {\sc Matlab} random number generator (Mersenne Twister with 19937 bits) and associated state initiation. The C++ implementation was rigorously analysed for any biases or implementation errors, until an exact replication of the original {\sc Matlab} implementation under both unit (single simulation step) and integration (outcome over a series of steps) compliance was achieved. Integration testing obtained an exact state match between 14,400 reference points between the {\sc Matlab} and C++ implementations (i.e. the model state every 10 steps) over a simulated 10 day tumour growth period.  From there, the C++/GPU accelerated articulation of the C++/CPU code was prepared. A series of randomly generated protocols was evaluated with both CPU and GPU simulations and the results were statistically compared. For details, c.f.~Fig.~\ref{fig:CPUvsGPU_normal} panel A together with ~Fig.~\ref{fig:CPUvsGPU_normal} panel B where p-values of the two-tailed t-test comparing corresponding GPU and CPU distributions are reported.

Finally, the normality of the GPU implementation was examined using the fast GPU implementation of the model. The final number of occupied grid sites was chosen as a representative measure of the tumour state because it is used to calculate the GA fitness function. To check if a data set can be well described by a normal distribution, a single irradiation protocol was randomly selected, and then 2\,000 evaluation of that protocol were run on the randomly selected tumour from the library. In Fig.~\ref{fig:CPUvsGPU_normal} (D)  a histogram of the results is plotted which, together with the Kolmogorov-Smirnov test (p-value = 0.38), indicate that the null-hypothesis, i.e. that the distribution is normal, cannot be rejected. In Fig.~\ref{fig:CPUvsGPU_normal} (C)  p-values of Kolmogorov-Smirnov test for 60 randomly generated protocols comparing GPU and CPU implementations are presented.

\subsection*{The Genetic Algorithm (GA)}

\subsubsection*{GA Overview}\label{sec:GA_approach}

\begin{figure}[ht]
    \centering
    \includegraphics[width=0.9\textwidth]{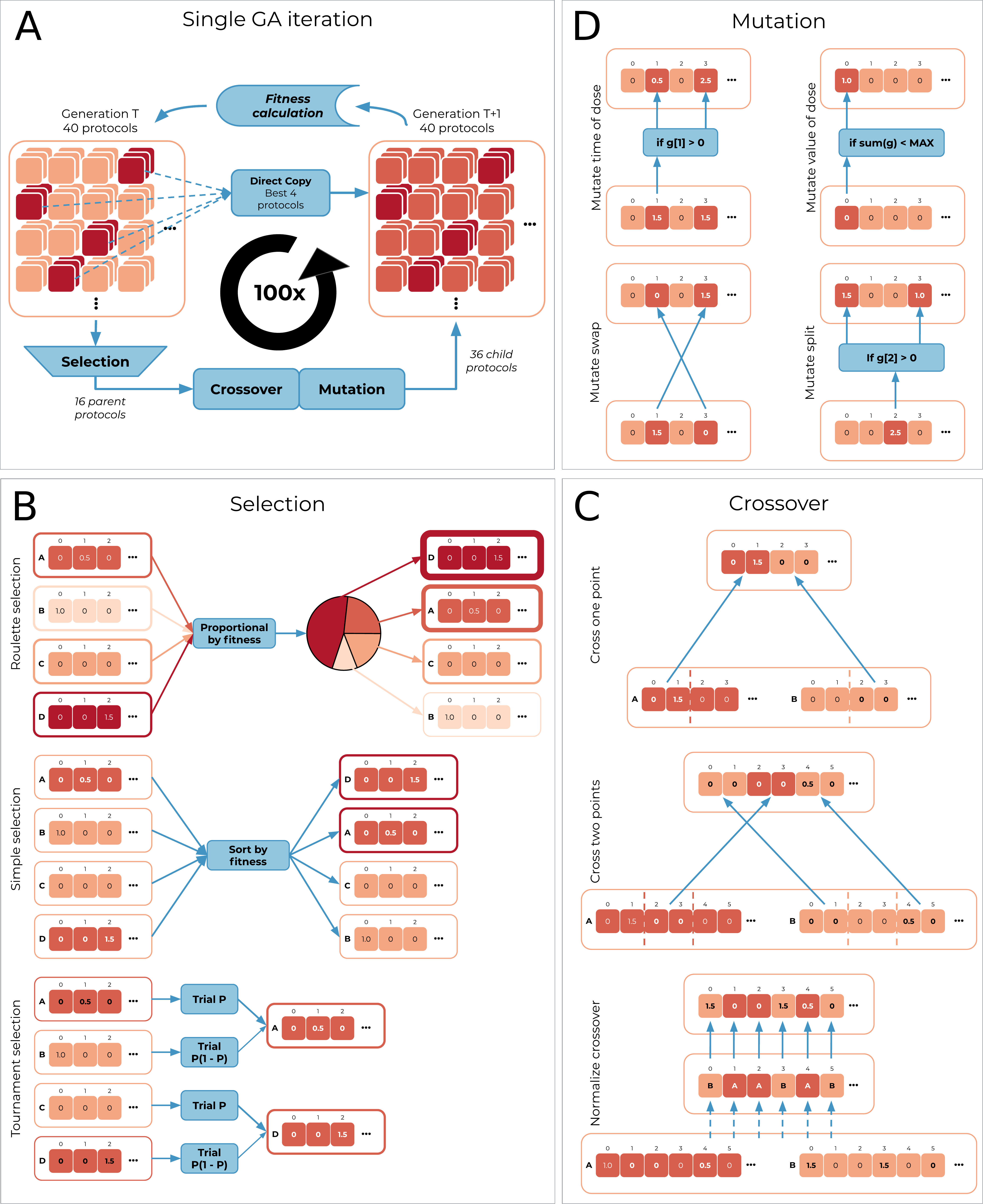}
    \caption{{\bf The Genetic Algorithm (GA) scheme.} A single generation consists of 40 candidate protocols. A new generation is formed by graduating 10\% (4) protocols directly without any alteration (A). A further 16 protocols are selected in one of three ways: Roulette (proportional chance by fitness), Simple (simple sort and take the top few), or Tournament (pair-wise competition on fitness). Next 36  protocols are formed from the 16 parents via applying the crossover (C) and mutation (D) operators.}
    \label{fig:GA_scheme}
\end{figure}

In order to find heuristically optimal protocols, we use genetic algorithms (GAs)~\cite{Holland:2012, goldberg} to explore the vast, combinatorial space of possible protocols. GAs are well suited to high dimensional, non-linear search, especially where quasi-optimal solutions are likely composed of modular sub-components.  In the GA approach irradiation protocols are encoded as genotypes represented as vectors. The quality (fitness) of genotypes is evaluated by a fitness function~\eqref{eq:final_fit_fun} for which the values are calculated using the main GPU implementation of the underlying computational model. The best genotypes are selected using a selection operator, and are later used to create new genotypes using crossover and mutation operators. This procedure is schematically presented in Fig.~\ref{fig:GA_scheme}. Together, the aim of the GA approach is to amplify the abundance of the most effective protocols, and protocol sub-modules, whilst retaining (and introducing) sufficient novelty to each generation to avoid the population becoming `stuck' within a locally-optimal (but perhaps not globally optimal) basin of attraction. Since each problem will have its own solution landscape, it is common practice to experiment with a variety of GA operators (see below) within the main selection, crossover, and mutation steps.

\subsubsection*{GA Resolution, constraints}
The CA model is designed with a resolution such that each hour between doses is split into 600 units of model time (a step). For our treatment protocols, we used a minimal step equal to 300 units, i.e. 30~min, given that smaller intervals between fractions may be difficult to deliver in the clinic. The full treatment time is set to 5~days (120~hours), which implies that each protocol should direct irradiation treatment over 240~time-steps. Additional constraints were established for possible fractional doses. First, following~\cite{Angus2014}, we assume that the sum of all dose values for a~single protocol should not be higher than 10~Gy to ensure that total delivered dose, affecting both the tumour and the healthy tissues, do not exceed a standard clinical practice for low-dose, multi-fraction, irradiation schemes \cite{Rosenstein:2004vs,ORourke:2009,BoardoftheFacultyofClinicalOncology:2006wq}. Moreover, the value of an acceptable single dose should be at a minimum 0.25~Gy and could be increased by 0.25~Gy per step, up to the maximal single dose ($d^{max}$), between 1.25~Gy and 4.5~Gy, depending on a particular experimental setting. In addition, to simulate realistic treatment protocols with doses given to the patient according to the scheduled intervals, we restricted the minimal time interval between doses to vary from 6 to 48 hours.

\subsubsection*{Protocol Representation}
In the implemented code, for GA, each protocol is represented by a sparse vector of 240 dose values, where each position corresponds to a specific point in time. Zeroed values in this representation indicate lack of dose at a corresponding time. A non-zero value in this representation indicates the fractional dose to deliver, at that time point. For example, a value of 1.25 in the fourth vector position would indicate that a 1.25~Gy fraction is delivered 2 hours (four times 30 minutes) after the initialisation of the treatment.


\subsubsection*{The Main GA Loop}


Each GA generation $T$ consists of 40 protocols which, after fitness determination in the full EMT6/Ro GPU computational model, is used as the genetic material for the formation of the subsequent generation, $T+1$ (c.f. Fig.~\ref{fig:GA_scheme}). In the initial step of the GA algorithm, 10\% (i.e., 4) of the best protocols in generation $T$ (according to simple selection) are \emph{copied} from generation $T$ into generation $T+1$ without modification (and retained in the genetic pool of generation $T$). Next, 40\% of the best protocols (i.e., 16) are selected from generation $T$ to become \emph{parents} using one of three \textit{selection} operators: \textit{roulette}, \textit{simple}, or \textit{tournament}. Note, \emph{selection} may cause one or more protocols from the initial copy step to be also present in the parent set. In a third step \emph{crossover} and \emph{mutation} operators are carried out in sequence to form 36 \emph{child} protocols. Crossover uses one of the available operators: \textit{single-point crossover}, \textit{two-point crossover}, or \textit{uniform crossover}. Mutation uses one of the available operators: \textit{swap mutation}, \textit{split mutation}, \textit{dose time mutation}, or \textit{dose value mutation}. Thus, the 36 generated \textit{child} protocols are combined together with the initial 4, copied protocols, to create the new generation $T+1$. Each protocol in the $T+1$ generation is evaluated using EMT6/Ro GPU computational model and \textit{fitness} function. After a prearranged number of iterations, representing a five-day treatment period, followed by a five-day re-growth period, the experiment is finished. The particular type of \textit{selection}, \textit{crossover} and \textit{mutation} operators used is set in the configuration of a given experiment.


\subsubsection*{The fitness function \& experimental setup}
\label{sec:fitness_function}

One of the crucial elements of our experiments is to determine the value of a fitness function $f$, required for GA as the fitness indicator. We use
\begin{equation}\label{eq:final_fit_fun}
   f(p_{i,j}) = 1\,500 - n_{i,j}, 
\end{equation}
where $p_{i,j}$ denotes the considered protocol $i$ evaluated on a single tumour $j$, 1\,500 is the maximal number of possibly occupied sites, $n_{i,j}$ denotes the number of the occupied sites after testing protocol $i$ on tumour $j$ (at day 10). In the performed experiments, we pass the treatment protocol into the CA model and obtain the fitness score. In order to assess the robustness of these results, each protocol is evaluated on 10 different tumours from the library using 2 GPUs, 4 times per GPUs per tumour. This allows us to calculate 80 samples for every single treatment protocol. Thus, the fitness for the particular tested protocol is the mean from the fitnesses of all its evaluations.

\subsubsection*{Calculating tumour suppression improvement}

To estimate the improvement of tumour suppression for a considered protocol $i$ ($ITS_i$) versus a reference protocol evaluated over 10 tumours we calculate the relative difference between the average number of occupied sites after testing protocol $i$ on 10 tumours ($\bar n_{i}$) and the average number of occupied sites for a reference protocol (tested on the same 10 tumours, $\bar n_{ref}$)
\begin{equation}\label{eq:supression}
    ITS_i=\frac{\bar n_{ref}-\bar n_{i}}{\bar n_{ref}}.
\end{equation}









\section*{Acknowledgements}


This work was supported by grant no.~2015/19/B/ST1/01163 of National Science Centre, Poland, ``Mathematical models and methods in description of tumour growth and its therapies''. This research was carried out with the support of the Interdisciplinary Centre for Mathematical and Computational Modelling at the University of Warsaw (ICM UW \cite{ICM}) under grants no G74-17 and GR79-29. Author contributions: All authors designed the study, WO, RB and PG carried out the experiments, all authors contributed to the analysis of the results, and to writing and editing the paper. The authors declare no competing interests. Code and data to reproduce the numerical experiments in this paper are available to reviewers at \cite{SourceCUDA}.

\section*{Supplementary Materials}
\label{appendix_a}
For Supplementary Materials see additional pdf file.







\section*{Figures}

\end{document}


\vspace*{0.2in}


{\Large
\textbf\newline{S1 Appendix for identifying promising candidate radiotherapy protocols via GPU-GA~\textit{in-silico} } 
}
\newline
\\
Wojciech Ozimek\textsuperscript{1},
Rafa\l{} Bana{\'s}\textsuperscript{2},
Pawe\l{} Gora\textsuperscript{3},
Simon D. Angus\textsuperscript{4,*},
Monika J. Piotrowska\textsuperscript{5},
\\
\bigskip
\textbf{1} Ardigen SA, 76 Podole Street, 30-394 Krakow, Poland
\\
\textbf{2} NVIDIA Corporation, Al. Chmielna 73, 00-801, 00-001 Warsaw, Poland
\\
\textbf{3} Institute of Informatics, University of Warsaw,
Banacha 2, 02-097 Warsaw, Poland
\\
\textbf{4} Department of Economics and SoDa Laboratories, Monash Business School, Monash University,\\ Wellington Rd, Clayton, 3800, Australia
\\
\textbf{5} Institute of Applied Mathematics and Mechanics, University of Warsaw, Banacha 2, 02-097 Warsaw, Poland
\bigskip

%
%





* simon.angus@monash.edu.au


\section*{Infrastructure used in experiments}
All the experiments with programs prepared in C++ were run using a computational infrastructure provided by the Interdisciplinary Centre for Mathematical and Computational Modelling of the University of Warsaw (ICM) (\textit{39}). Experiments requiring CPUs were run on $36$ cores of a Rysy cluster, with Intel(R) Xeon(R) Gold 6154/6252 CPU processors (3.7~GHz) and 380 GB RAM. Experiments requiring GPUs were run on 4~GPUs of graphical processing units NVIDIA V100 on the same cluster - Rysy. At early stage of the project, some preliminary tests were also run on the Topola cluster.

\section*{GA Operators: additional details}

\subsection*{Selection}
Selection procedure is an important step for GA to obtain the convergence and desired results. During the process, a portion of the current (parent) population is selected and used as a parental population for a new generation. It ensures diversity of protocols, enhances exploration of the search space, helps in preventing overfitting to relatively good protocols, and leads to higher possibility for finding better results. To empower results, we use 3~different selection algorithms, commonly applied in GA problems: \textit{simple selection}, \textit{roulette wheel selection} and \textit{tournament selection}, c.f. Fig.~4 (panel B) in the main article. 

The first method, \textit{simple selection}, sorts protocols by associated fitness values in descending order and selects 40\%  of them. This algorithm is very simple, however, it might limit the searching space for new possible treatment protocols and result in being stuck in local minima. 

To prevent such situation, we use two alternative algorithms, which might mitigate the aforementioned risk. In the \textit{roulette wheel selection} algorithm, called also fitness proportionate selection, we associate fitness values with the probability of selection. The probability $p_i$ for each protocol is calculated as 
\[
p_i = \frac{f_i}{\sum^N_{j=1}f_j},
\]
where $f_i$ denotes a fitness value for $i$-th protocol and $N$ is the number of treatment protocols in the considered generation ($40$). Next, we calculate the cumulative sum of those probabilities, summing them all to 1. To illustrate \textit{roulette wheel selection}, we can use a roulette wheel, where the proportions of the wheel segments are represented by probability $p_i$. We order the segments according to their order on the wheel and for each segment, we calculate sum of probabilities of all previous segments. For choosing a single treatment protocol, we take a random value $q$ between 0 and 1 and select a protocol associated with the first segment for which the partial sum of probabilities is higher or equal to $q$. We run the \textit{roulette wheel selection} algorithm in the loop to obtain treatment protocols as parents for the next generation (40\% i.e. 16 protocols). This method promotes protocols with higher probabilities, however, it allows to select weaker protocols with a small chance. This non-zero chance for weak protocols is an advantage over the simple selection which may help in escaping from the local minima.

For a similar purpose, we use the third selection algorithm, the \textit{tournament selection}, where we randomly select $k$~protocols and run tournaments among them, so that the best from the selected $k$~protocols wins the tournament and is eventually selected for the further consideration (crossover and mutation). For each protocol, the probability of being selected to a given tournament is associated with its fitness value - first, all protocols are sorted according to the fitness value, then the probability of selecting the $k$-th best protocol to the tournament is $p_i = q*(1-q)^{(k-1)}$ with a chosen probability $q$.


In our case, we repeat selection of $k=16$~protocols for a tournament $16$ times to obtain the desired number of parents protocols for crossover. Similarly to the \textit{roulette wheel selection}, this algorithm also allows weaker protocols to be selected. Comparing to the roulette selection, we can adjust the probability of selection by modifying the value of the parameter $q$. In this work we use the $q=0.9$ as it enables the \textit{roulette wheel selection} to select some worse performing protocols and reduces the risk of getting stuck in local minima.


In addition, for each GA generation we use simple selection to retain 10\% of the best protocols (according to their fitness score). These protocols are not taken into account in crossover and mutation operations, to prevent the loss of the best protocols, and copied directly to the next GA generation.

\subsection*{Crossover}
\label{crossover}
The goal of the crossover operation is to produce the new generation from the selected parents. In our experiments, the newly generated child inherits genes (doses at time-steps) from two randomly selected parents. This operation is repeated until the size of new population is appropriate. We introduce three different types of crossover algorithms: \textit{single-point crossover}, \textit{two-point crossover} and \textit{uniform crossover}, see Fig.~4 panel C in the main article. For crossover operator, we benefit from storing protocols as sparse vectors (with zeroes corresponding to the lack of dose and non-zeroes to a dose given at the particular time-step). This makes algorithms more conceptually straightforward and very similar to those presented in the literature (\textit{37}, \textit{38}).

In the \textit{single-point crossover}, we randomly choose two parent protocols (from the pool of previously selected patents) and select the same crossover point for both of them. Next, we swap the parts of protocols after the crossing points and obtain two children protocols. Crossing point is selected randomly, although with restricted range of choice between $25\%$ and $75\%$ of the length of a protocol. This limitation is important in the case of sparse representation, because in a situation when only a few first or last positions are swapped, the protocols will remain unchanged.

Second algorithm, \textit{two-point crossover}, is a variation of the aforementioned operator. Instead of selecting a single crossing point, we randomly selects two points. For one child protocol, the first and the last parts come from the first parent while the middle from the second parent. Analogously, the second child protocol has the middle part from the first parent and the first and the last parts from the second parent. In this case, we also introduce limitations for the location of crossover points positions. The range for the first point is set to between $25\%$ and $50\%$ of the length of a protocol, while for the second: between $50\%$ and $75\%$. Again the procedure is repeated as long as we get the required number of children protocols. 

The last introduced crossover operation is the so-called \textit{uniform crossover}, which also takes two parent protocols and returns two children protocols. However, the child protocol is created by selecting single positions, one-by-one, from both parent protocols with equal probability. If one child receives a gene from the first parent, then the other child receives a corresponding gene from the second parent on this position.

Clearly, protocols generated by \textit{crossover operators} might do not meet the constraints regarding the minimal time interval between doses and maximal single dose value and the total dose. To prevent the former issue, we designed  \textit{guard algorithms}. One of them identifies doses that violate time constraint and tries to assign them at another time-steps. If all the non-zero doses are too close to each other to allocate a dose between them, the algorithm skips dose allocation. The latter problem with exceeding the total acceptable dose  (10~Gy)  was resolved by another guard algorithm -- if the total dose exceeds 10~Gy then the algorithm starts to subtract the value of 0.25~Gy from the consecutive largest doses in a loop, until the sum of all doses for a protocol is acceptable.

\subsection*{Mutation}
Mutation is an essential genetic operator to sustain diversity of a population and to avoid a stuck of the GA in local minima. Mutation operators are applied with a low probability to reduce the possibility of random search and loss of properties of good protocols. On the other hand, usage of too low mutation probability  lower exploratory abilities of GA. In the presented study, we implement four different mutation operators: \textit{swap mutation}, \textit{split mutation}, \textit{dose time mutation} and \textit{dose value mutation}. For simplicity, in the case of mutation operators we use sparse vector representation of protocols, similarly as in the case of crossover operators.

\textit{Swap mutation} operator swaps two doses. We start by iterating over protocols and selecting (with a probability $p$) a random position on the chromosome to swap. Such operation might break the constraints regarding minimal time interval between two doses and thus we apply the same \textit{guard algorithm} 
to overcome that issue.

\textit{Split mutation} operator randomly (with probability $p$) selects non-zero (larger than 0.25~Gy) dose and splits it into two smaller ones.
Next, the algorithm removes the split dose and allocates the smaller doses at randomly selected positions. It does not replace doses but appends them to existing values, only if the new, combined dose, is not larger than a maximal allowed single dose value (chosen for the particular GA experiment). If the new combined dose is larger than maximal allowed single dose, algorithm skips the operation and tries to find a better position for a dose. Similarly as in the case of the swap mutation, we used the \textit{guard algorithms} to meet the constraints of both minimal time interval between doses and the maximal value of a total dose.

The third operator, \textit{dose time mutation}, alters the position of doses in a protocol. With a probability of $p$, it randomly selects the dose at a source position from a protocol and the new destination position for it. Next,  source dose is added to the dose at the destination position and the dose at the source position is removed (zeroed). To meet the constraints, we clamp the dose at the destination position to maximal allowed single dose value. The remained dose is allocated back at the source position. Using this algorithm, we avoid losing dose values from a protocol keeping the sum of doses for the protocol at the same level.

The last introduced operator, \textit{dose value mutation}, changes the value of selected dose. With a probability $p$ we allocate a new dose to the protocol. First, we calculate the difference between a sum of dose values and the maximum total dose (10~Gy). If it is smaller or equal to zero we do nothing, otherwise (which may happen only in the case of applying more than $1$ mutation operator to a given protocol), we randomly select an existing dose in the protocol and replace it with a larger dose, which does not break the maximum allowed sum limit. However, this might break a time interval constraints, so again we use the \textit{guard algorithm}. The GA algorithm allows for multiple types of mutation and also repeated use of a given type of mutation to be applied for a single protocol. Such a combination of mutations induces the possibility for a loss in a total dose. The benefit of this mutation operation is the increase of the sum of doses for a protocol, alleviating the possible loss of dose values from split and swap mutations.

\section*{GA metrics}
For every experiment we collected data to further evaluate the performance of protocols. For each GA loop iteration we stored the best fitness function values and associate the best protocols. Next, we use them to analyse and compare the best protocols from different iterations and experiments. In order to check whether our GA model does not overfit, for each iteration we store the average value of a fitness function for all protocols.

\section*{GA experiments}
Due to the limited resources and relatively long computation time, we organised the experiments in batches. Each batch was focusing on particular assumptions and hypotheses from previous runs, so results of experiments from previous batches influenced our decisions on how to configure the next series of experiments. 

In total, we ran $8$ series of experiments, their settings are presented in 
S1~Table. Below, we focus on the last batch among all the performed experiments since others served for the improvement of used approaches by testing various settings and configurations of GA algorithms.

Based on the results of previous batches, in the eighth batch we decided to run the best performing configuration of operators for all available single maximum doses and time-delay intervals. We conducted experiments with the three introduced selection operators, \textit{roulette wheel selection}, \textit{tournament selection} and \textit{simple selection}. We used \emph{two-point crossover} operator and a combination of three mutation operators, namely: \textit{swap mutation}, \textit{split mutation}, and \textit{dose time mutation}. To capture all of the possible time intervals and maximal single dose configurations, we finally decided to use intervals of 6, 12, 24, and 48~h and single admission doses limited to 1.25, 2.0, 2.5, 3.0, 3.5, 4.0 and 4.5~Gy. This gave us a total of 84 configurations ($3$ selection operators $\times$ $4$ time delay minimal intervals $\times$ $7$ maximal single dose values), increasing a searching space compared to the previous study (\textit{7}). Each individual configuration was run three times to ensure robustness of the results. Because our GA algorithm usually converged after 30 iterations we decide to use 50 iterations of GA per single run. During evaluation of results, we decided to conduct additional 3 runs up to 100 iterations of GA for \textit{roulette wheel selection}, because for some experiments, the variability of fitness function was higher. However, it did not affect the results. Each protocol was evaluated by simulations 80 times (8 times for each of 10 tumours in library, see Section \textit{The fitness function \& experimental setup} in the main article), the average and standard deviations of evaluation values were calculated. For each configuration, we collected values for 3 best protocols from all iterations obtained during those 10 GA runs.

The results of these experiments for 3 best protocols obtained during GA runs are summarised in 
S2~Table. The superiority of \emph{simple selection} over \emph{tournament} and \emph{roulette wheel} is visible. Clearly, the results for \emph{simple selection} are dominant for the shortest time interval of 6~h.
